\documentclass{aa}
\usepackage{graphicx,natbib}
\bibpunct{(}{)}{;}{a}{}{,}

\def\apj{ApJ}

\def\flux{erg s$^{-1}$ cm$^{-2}$}
\def\lum{erg s$^{-1}$}

\begin{document}

\sloppypar

%
   \title{Origin of the Galactic ridge X-ray emission}

   \author{M.~Revnivtsev \inst{1,2} \and S.~Sazonov \inst{1,2} \and
   M.~Gilfanov \inst{1,2} \and E.~Churazov \inst{1,2} \and R.~Sunyaev\inst{1,2}}

   \offprints{mikej@mpa-garching.mpg.de}

   \institute{
              Max-Planck-Institute f\"ur Astrophysik,
              Karl-Schwarzschild-Str. 1, D-85740 Garching bei M\"unchen,
              Germany,
      \and
              Space Research Institute, Russian Academy of Sciences,
              Profsoyuznaya 84/32, 117997 Moscow, Russia
            }
  \date{}

        \authorrunning{Revnivtsev et al.}

   \abstract{We analyze a map of the Galactic ridge
   X-ray emission (GRXE) constructed in the 3--20~keV energy band from
   RXTE/PCA scan and slew observations. We show that the
   GRXE intensity closely follows the Galactic near-infrared surface
   brightness and thus traces the Galactic stellar mass
   distribution. The GRXE
   consists of two spatial components which can be identified with the
   bulge/bar and the disk of the Galaxy. The parameters of these
   components determined from X-ray data are compatible with those derived
   from near-infrared data. The inferred ratio of X-ray to near-infrared
   surface brightness $I_{\rm 3-20 keV}$($10^{-11}$ \flux\ deg$^{-2}$)/$I_{\rm
   3.5\mu m}$(MJy/sr)=$0.26\pm0.05$,  and the ratio of X-ray to
   near-infrared luminosity $L_{\rm 3-20 keV}/L_{\rm 3-4 \mu m}=
   (4.1\pm0.3)\times10^{-5}$. The corresponding ratio of the 3--20~keV
   luminosity to the stellar mass is $L_{\rm x}/M_{\odot}=
   (3.5\pm0.5) \times 10^{27}$ erg s$^{-1}$, which agrees within
   the uncertainties with the cumulative emissivity per unit stellar
   mass of point X-ray sources in the Solar neighborhood, determined in an
   accompanying paper (Sazonov et al.). This suggests that the bulk of
   the GRXE is composed of weak X-ray sources, mostly cataclysmic variables and
   coronally active binaries.  The fractional contributions of
   these classes  of sources to the total X-ray emissivity determined from the
   Solar neighborhood data can also explain the GRXE energy
   spectrum. Based on the luminosity function of local X-ray sources
   we predict that in order to resolve 90\% of the GRXE into discrete
   sources a sensitivity limit of $\sim10^{-16}$ \flux\ (2--10~keV)
   will need to be reached in future observations.
   }

   \maketitle

%

\section{Introduction}
There are two major large-scale extended features in the X-ray
sky (above 2 keV): the almost uniform cosmic X-ray background (CXB,
\citealt{giacconi62}) and an emisison concentrated toward the
Galactic plane --  the Galactic ridge X-ray emission (GRXE, see e.g.
\citealt{worrall82}). While over the last two decades it has been
firmly established that the CXB is a superposition of a large number
of discrete extragalactic sources (namely active galactic nuclei,
see e.g. \citealt{giacconi02}), the origin of the GRXE remains
unexplained.

Exploration of the GRXE by different observatories has revealed that
it is concentrated near the inner Galactic disk, extending tens of
degrees in longitude and a few degrees in latitude
\citep{cooke70,worrall82,warwick85, warwick88,yamauchi90}, and likely
has a central bulge-like component \citep{yamauchi93,mikej03}. The energy
spectrum of the GRXE contains a number of emission lines of highly ionized
heavy elements, indicating that the emission should be
thermal with a temperature of up to 5--10 keV
\citep{koyama86,koyama89,tanaka02,muno04}. The total GRXE luminosity
has been estimated at $\sim$1--2$\times 10^{38}$ erg s$^{-1}$
\citep{yamauchi93,valinia98}.

\begin{figure*}
\includegraphics[height=\textwidth,angle=-90]{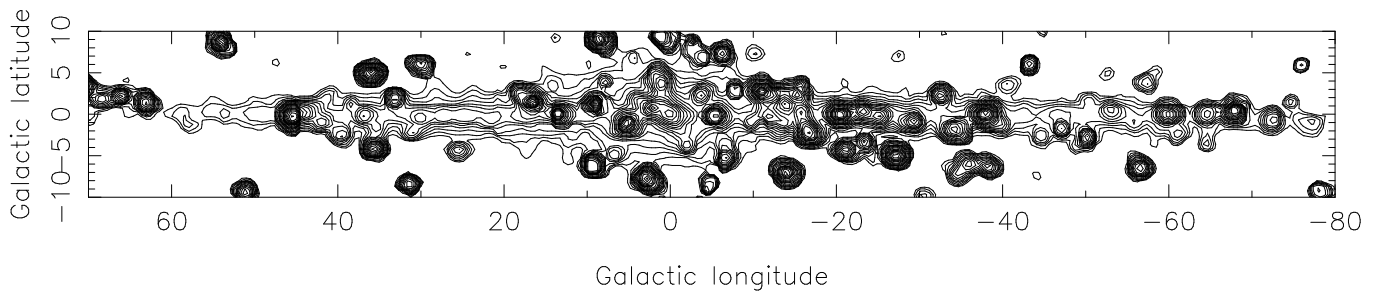}
\caption{RXTE/PCA map of the sky around the Galactic plane in the
energy band 3--20 keV. Contour levels are logarithmically spaced
with a factor of 1.4, with the lowest contour corresponding to an
intensity of $10^{-11}$ \flux\ deg$^{-2}$. Map clearly exposes many
bright point sources and an underlying unresolved
emission\label{map}}
\end{figure*}

The GRXE has been detected at least up to 20--25 keV energies
\citep{valinia98,mikej03}, and its spectrum in the 3--20 keV range
consists of a continuum, which can be approximated by a power law of
photon index $\Gamma\sim2.1$, and powerful lines at 6--7 keV
energies. Also a detection of GRXE at energies $>$40 keV (at
Galactic longitude $l=95$) was reported (e.g.
\citealt{skibo97,valinia98}), but it now appears that those
CGRO/OSSE measurements were strongly contaminated by a few
unresolved sources, including active galactic nucleus IGR
J21247+5058 recently discovered by the INTEGRAL observatory
\citep{masetti04}. The IBIS telescope aboard INTEGRAL, capable of
resolving point-like sources with flux $>$few$\times 10^{-11}$
\flux\ in crouded regions, has not detected the GRXE at energies
above $\sim$40 keV \citep{lebrun04,terrier04}.

Soon after discovery of the GRXE it was proposed that it might
consist of a large number of weak Galactic X-ray point sources, e.g.
quiescent low-mass and high-mass X-ray binaries, cataclysmic variables,
coronally active binaries etc
\citep{worrall82,worrall83,koyama86,ottmann92,mukai93}. However, it
was not possible to draw a solid conclusion due to lack of detailed
information about the space densities and X-ray luminosity
distributions of these classes of X-ray sources.

 Unless the GRXE is truly diffuse emission, it should eventually be
possible to resolve it into a finite number of discrete sources. As
the sensitivity of X-ray telescopes has been increasing, a
progressively higher fraction of the GRXE has been resolved
\citep{worrall82,warwick85,
  sugizaki01}. However, even the deepest observations of Galactic plane
regions by the currently operating Chandra and XMM-Newton
observatories, in which point-source detection sensitivities $F_{\rm
x}>3\times 10^{-15}$ \flux\ \citep{ebisawa01,hands04,ebisawa05} were
achieved, resolved not more than 10--15\% of the GRXE. This was
regarded as a strong indication of the GRXE being truly diffuse.

However, the hypothesis of diffuse origin of the GRXE meets strong
difficulties \cite[e.g.][]{koyama86,sunyaev93,tanaka99,tanaka02}.
The main problem is that the apparently thermal spectrum of the GRXE
implies that the emitting plasma is so hot ($\sim $5--10 keV) that
it should be outflowing from the Galactic plane. A large energy
supply is then required to constantly replenish the outflowing
plasma.

Resolving the GRXE is additionally complicated by the fact that at fluxes
near the present-day sensitivity limit
($\sim 10^{-15}$ \flux) extragalactic sources outnumber Galactic ones even in
the Galactic plane (e.g. \citealt{ebisawa05}). Since identification of
weak sources detected in deep X-ray surveys of the Galactic plane is
usually problematic and the CXB varies significantly on
sub-degree angular scales, so far it has only been possible to place
upper limits on the fraction of the GRXE resolved into Galactic X-ray
sources (as mentioned above).

The only place where Galactic sources dominate over extragalactic ones is the
inner 10\arcmin\, of the Galaxy \citep{muno03}, and Chandra has
resolved up to 30\% of the ``hard diffuse''
emission in certain parts of this region \cite[region ``Close'' in this
  paper]{muno04}. Moreover, the flux-number distribution of
Galactic X-ray sources detected in this Galactic Center survey shows no cutoff
down to the Chandra detection limit, implying that at least an order of
magnitude deeper observations will be needed to resolve the bulk of the hard
X-ray emission from the Galactic Center if the source flux-number distribution
continues with the same slope to lower fluxes than presently
accessible.

It is therefore worth considering alternative ways to solve the
problem of the GRXE origin, in particular via studying its spatial
distribution. The distribution of the GRXE over the sky is still
poorly known, mainly because of its large extent (approximately
$120^\circ \times10^\circ$) and low surface brightness
($<$few$\times 10^{-11}$ \flux\ deg$^{-2}$). The investigation of
the GRXE in the HEAO-1/A2 experiment
\citep[e.g.][]{iwan82,worrall82} was significantly hampered by point
source confusion, whereas instruments with much better spatial
resolution were not able to cover a sufficiently large solid angle of the
sky \citep[e.g.][]{sugizaki01,hands04}.

In this paper we present a brightness distribution of the GRXE measured in the
3--20 keV energy band by the RXTE/PCA instrument and show that
this distribution closely follows the near-infrared brightness of the Galaxy,
known to be a good tracer of the stellar mass
distribution. We further compare the inferred Galactic ridge X-ray
emissivity per unit stellar mass with the cumulative emissivity of
point X-ray sources in the Solar neighborhood, determined by
\cite{sazonov05}, and argue that the bulk of the GRXE is likely
composed of weak discrete sources of known types.

\section{RXTE observations and analysis}

The best instrument so far for large-scale mapping of the X-ray
(above 2~keV) sky is the PCA spectrometer aboard the RXTE
observatory \citep{rxte}. It combines a large effective area ($\sim
6400$ cm$^{2}$ at 6 keV) and a moderate field of view ($\sim1^\circ$
radius). The latter enables both to achieve a good coverage of the
sky in the course of the mission (as compared to focusing
instruments) and to alleviate the source confusion problem (as
compared to X-ray collimators with larger fields of view). Over its
10-year lifetime RXTE/PCA has performed a large number of scan and
slew observations over the whole sky, and we previously made use of
these observations to catalog X-ray sources detected at high
Galactic latitudes \citep{mikej_xss,sazonov04} and to study the CXB
\citep{mikej_cxb}. We now take advantage of the good coverage of the
Galactic plane region by RXTE/PCA to study the GRXE.

We use the same set of observations and apply the same analysis as in
Revnivtsev et al. (2004). Throughout the paper we will use only data
from the first layers of all PCA detectors (PCUs) in the 3--20 keV energy
band. The net exposure time of the utilized observations normalized to the
single PCU effective area ($\sim$1300 cm$^2$) is approximately 29 Ms.

\subsection{Flux measurement uncertainties}

Apart from statistical uncertainties (photon noise), PCA
measurements of X-ray flux are subject to systematical
uncertainties. The current version of the RXTE/PCA software
(LHEASOFT 5.2) enables an accuracy of background (instrumental plus
the CXB flux averaged over the whole sky) subtraction of $\sim
0.02-0.03$ cnts/s/PCU/beam in the 3--20 keV energy band, which
corresponds to  $\sim 2\times 10^{-13}$ \flux\ deg$^{-2}$ for a
Crab-like spectrum \citep[e.g.][]{craig02}.

 Another factor that needs to be taken into account in studying the
GRXE (with subtracted contribution of bright point sources) is
small-scale variations of the CXB intensity (mainly due to
unresolved extragalactic sources within a 1-deg$^2$ field of view of
the PCA), which introduces an $rms$ uncertainty of $1.5\times
10^{-12}$ \flux\ deg$^{-2}$ (3--20~keV) in measuring the X-ray
surface brightness in a given direction \citep{mikej_cxb}. Note
however that for most of the sky regions studied here this
uncertainty does not exceed statistical errors.

Almost everywhere in the Galaxy, except in the very central regions
where we practically cannot study the GRXE because of high
concentration of bright point sources, the interstellar gas column
density does not exceed $\sim 2-3\times 10^{22}$ atoms
cm$^{-2}$. Photoabsorption in this gas attenuates the
GRXE flux at 3--20 keV by less than 5\%. We neglect this
small effect in the subsequent analysis.

\section{Map of the GRXE}

In Fig.~\ref{map} we present an X-ray intensity map of the sky around
the Galactic plane convolved with the response of the PCA collimator
(triangular shape with a radius $\sim1^\circ$, see
\citealt{mikej_cxb}). The contour levels on this map are
logarithmically spaced with a factor of 1.4, and the lowest shown level
corresponds to an X-ray intensity $\sim 10^{-11}$ \flux\ deg$^{-2}$.

The map clearly exposes many bright point-like sources and an underlying
unresolved emission -- the GRXE. Henceforth we reserve the term "GRXE"
to describe Galactic X-ray emission which cannot be
resolved into discrete sources with flux higher than $10^{-12}$
\flux. We note that the exact value of the limiting flux is
unimportant unless it is higher than $\sim 10^{-11-10.5}$ \flux\
\citep{sugizaki01,hands04}. This is explained by the fact that the density of
sources with $F_{\rm x}>10^{-12}$ \flux\ is less than 1 deg$^{-2}$
almost everywhere except in the central degree of the Galaxy.

The GRXE was given its name because it was originally detected as a
prominent narrow ($\sim$ 1--2$^\circ$) band of unresolved emission
along the Galactic plane \citep[e.g.][]{bleach72,worrall82}. However,
it has since then become more evident that the unresolved X-ray emission of
the Galaxy contains both a disk-like and a bulge-like component
\citep{yamauchi93,mikej03}. The exponential scale height of the disk
component of the GRXE is $\sim 1.5^\circ$
\citep{worrall82,warwick85,yamauchi93}, whereas it is much larger for the
Galactic bulge -- up to 3--5$^\circ$\citep{yamauchi93, mikej03}. Both
components can now be clearly seen on the RXTE map shown in Fig.~\ref{map}.

We will now separately consider the bulge and disk components of the GRXE.

\subsection{Galactic bulge}

The observed intensity distribution of the GRXE in the Galactic
Center region is strongly affected by the bright point sources
located there. Therefore, in order to study the underlying GRXE we
should  mask out bright point sources. To this end we filtered out
$1.5^\circ$-radius regions around point sources with flux higher
then $\sim 1$ cnts/s/PCU/beam$\sim 1.2\times 10^{-11}$ \flux\ (this flux
conversion corresponds to the measured GRXE spectrum, \citealt{mikej03}). The
catalog of detected point sources will be published elsewhere. This
flux limit corresponds to a source luminosity $L_{\rm x}\sim
10^{35}$ erg s$^{-1}$ for a Galactic Center distance of 8.5 kpc.

Since the number density of sources with flux higher than $\sim 1$
cnts/s/PCU/beam is quite high, we have practically no data left at
$|b|<1-2^\circ$ upon applying the above filtering procedure
(similar to the analysis of Revnivtsev 2003). At higher latitudes
the number density of bright point sources drops significantly
\cite[e.g.][]{grimm02} but the applied mask still severely reduces
the coverage of the bulge.

Relatively bright sources ($L_{\rm x}>10^{34}$ erg s$^{-1}$, or
$F_{\rm x}>10^{-12}$ \flux) below our filtering threshold might lead
to significant deviations of the observed X-ray intensity map
from the actual GRXE brightness distribution. However, as was
already mentioned, the density of such sources is less than $\sim$1
deg$^{-2}$ (except in the central degree of the Galaxy), while the
typical intensity of the GRXE is $\sim 10^{-11}$ \flux\, deg$^{-2}$
\cite[e.g.][]{sugizaki01}. We thus expect that our filtering
criterion enables recovering the surface brightness distribution of
the GRXE with approximately 10\% accuracy.

\begin{figure}
\includegraphics[width=\columnwidth,bb=50 180 570 550,clip]{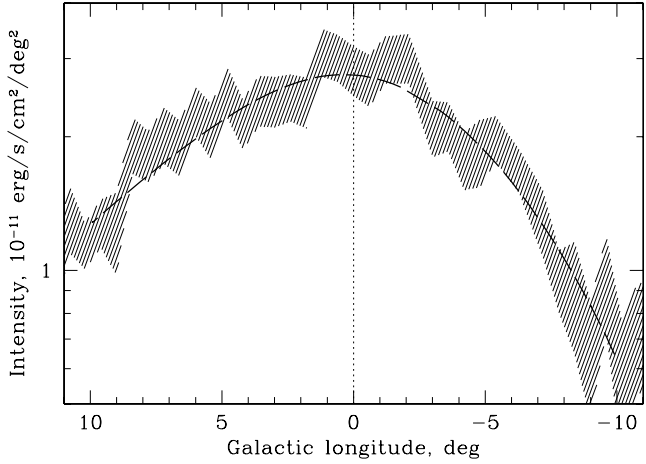}
\caption{Intensity profile of the bulge/bar component of the GRXE in
the slice $3.0^\circ<|b|<3.5^\circ$ parallel to the Galactic plane. The shaded
region indicates the measurement uncertainty including the 10\%
systematics described in the text. The dashed line is the best-fit
model of the bulge/bar defined by parameters given in Table~\ref{pars}.}
\label{bulge_slice}
\end{figure}

\begin{figure}
\includegraphics[width=\columnwidth,bb=50 180 570 580,clip]{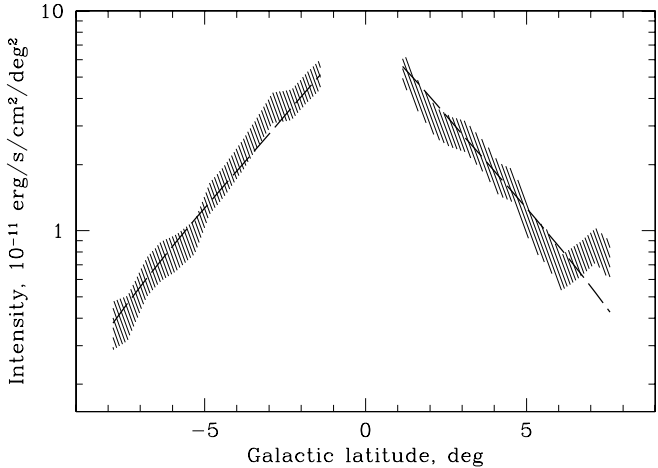}
\caption{Intensity profile of the bulge/bar component of the GRXE
perpendicular to the Galactic plane at $|l|<4^\circ$. The shaded
region indicates the measurement uncertainty including the 10\%
systematics. The dashed line is the best-fit model defined by
parameters given in Table~\ref{pars}.}
\label{bulge_slice_vert}
\end{figure}

In order to construct an intensity profile of the bulge component of
the GRXE along the Galactic plane we selected latitudes
$3.0^\circ<|b|<3.5^\circ$, where the disk component of the GRXE is
weak while the bulge component is still relatively bright. Also an
intensity profile of the bulge perpendicular to the Galactic plane was
constructed from observations at $|l|<4^\circ$ excluding $|b|<1^\circ$.
The resulting profiles are shown in Fig.~\ref{bulge_slice} and
Fig.~\ref{bulge_slice_vert}, respectively. The shaded regions indicate the
measurement uncertainties including the 10\% systematics
described above. One can see that the GRXE intensity profile
is relatively symmetric with respect to the Galactic plane but is
not symmetric along the plane with respect to $l=0$.

It is natural to compare the inferred morphology of the GRXE bulge
with the Galactic stellar bulge/bar
\citep{bahcall80,blitz91,weiland94,dwek95}. For this purpose we use
the analytic model of the bulge/bar stellar volume emissivity
developed by \cite{dwek95}, namely model G3 in this paper. This
model is derived from the near-infrared surface brightness
distribution of the Galaxy measured by COBE/DIRBE, which is believed
to be a very good tracer of the stellar mass distribution.

Let us parameterize the volume emissivity of the bulge component
of the GRXE similarly as was done for the stellar bar by
\cite{dwek95}. The observed intensity of the GRXE is proportional
to the line-of-sight integral of the bulge/bar volume emissivity
predicted by the model. Specifically, the X-ray flux $F(l,b)$ measured
within a solid angle $d\Omega$ is given by

$$
F(l,b)={d\Omega\over{4\pi}}\int_{0}^{\infty} \rho(x,y,z) ds.
$$

The volume emissivity of the bar is given by
$$
\rho_{\rm bulge} (x,y,z)=\rho_{0,\rm bulge} r^{-1.8} \exp(-r^3),
$$

where
$$
r=\left[\left({x\over{x_0}}\right)^2+\left({y\over{y_0}}\right)^2+\left({z\over{z_0}}\right)^2\right]^{1/2}.
$$

The major axes $x$ and $y$ of the bar are assumed to lie in the Galactic plane
and the $x$ axis is inclined by an angle $\alpha$ to our line of sight.

Just by varying the normalization of the G3 model of \cite{dwek95} we
obtain fairly good fits to the measured GRXE intensity
profiles, with reduced $\chi^2 \sim 1.3$. This implies that the
longitudinal asymmetry clearly evident in Fig.~\ref{bulge_slice} is a
natural observational consequence of the (triaxial ellipsoid) bar
inclined to our line of sight. The near end of the bar lies in the first
Galactic quadrant.

\begin{table}
\caption{Best-fit parameters of the bulge/bar and disk components
 of the GRXE. The quoted errors on the values are 90\% statistical
 uncertainties.}
 \label{pars} \tabcolsep=5mm
\begin{tabular}{lc}
Parameter& Value\\
\hline
\multicolumn{2}{c}{Bulge/bar}\\
\hline
$\alpha$, $^\circ$ & $29\pm6$\\
$x_0$,kpc    &$3.4\pm0.6$ \\
$y_0$,kpc    &$1.2\pm0.3$\\
$z_0$,kpc    &$1.12\pm0.04$\\
$L_{\rm x,bulge}$, $10^{37}$ ergs s$^{-1}$  &$3.9\pm0.5^*$\\
$\chi^2$/dof&1.24\\
\hline
\multicolumn{2}{c}{Disk}\\
\hline
$R_{\rm disk}$   & 2.5(fixed) \\
$z_{\rm Sun}$,pc & $19.5\pm 6.5$\\
$z_{\rm disk}$,kpc & $0.13\pm0.02$\\
$L_{\rm x, disk}$, $10^{37}$ ergs s$^{-1}$ & $\sim 10^*$\\
$\chi^2$/dof& 0.93\\
\hline
\end{tabular}
\begin{list}{}
\item $^*$ -- without bright point sources
\end{list}

\end{table}

Allowing variations of the parameters we fitted the measured GRXE
intensity profiles shown in Fig.~\ref{bulge_slice} and
Fig.~\ref{bulge_slice_vert} in the ranges $|l|<10^\circ$ and
$1^\circ<|b|<8^\circ$, respectively. The derived best-fit parameters
are presented in Table~\ref{pars}. These values agree very well with
the parameters of the stellar bar obtained by \cite{dwek95}. We show
the best-fit longitudinal and latitudinal intensity profiles in
Fig.~\ref{bulge_slice} and Fig.~\ref{bulge_slice_vert}.

It is necessary to note that the determined value of the total X-ray
luminosity of the bulge given in Table~\ref{pars} should be regarded
with caution because our analysis did not involve detailed modelling
of the innermost (cuspy) region of the bulge.

\subsection{Galactic disk}

The severe confusion problem in the Galactic plane region prevents us
from constructing an intensity profile of the GRXE along the Galactic
plane with subtracted contribution of bright point-like sources. We
can however construct a combined intensity distribution of the GRXE and bright
point sources along the Galactic plane (within $|b|<0.5^\circ$), as
shown in Fig.~\ref{slice_plane}. On this profile the contributions of
discrete  sources are seen as rapid (on angular scales $\sim 1^\circ$)
variations of the intensity. The GRXE manifests itself as an
underlying smooth component. The statistical uncertainty does not
exceed $10^{-12}$ \flux\ deg$^{-2}$ almost everywhere in the Galactic plane.

Despite the strong contamination by bright discrete
sources we can obtain important information about the properties
of the GRXE disk from the longitudinal intensity profile shown in
Fig.~\ref{slice_plane}.

\subsubsection{Longitudinal extent}

\begin{figure}
\includegraphics[width=\columnwidth,bb=50 180 570 500,clip]{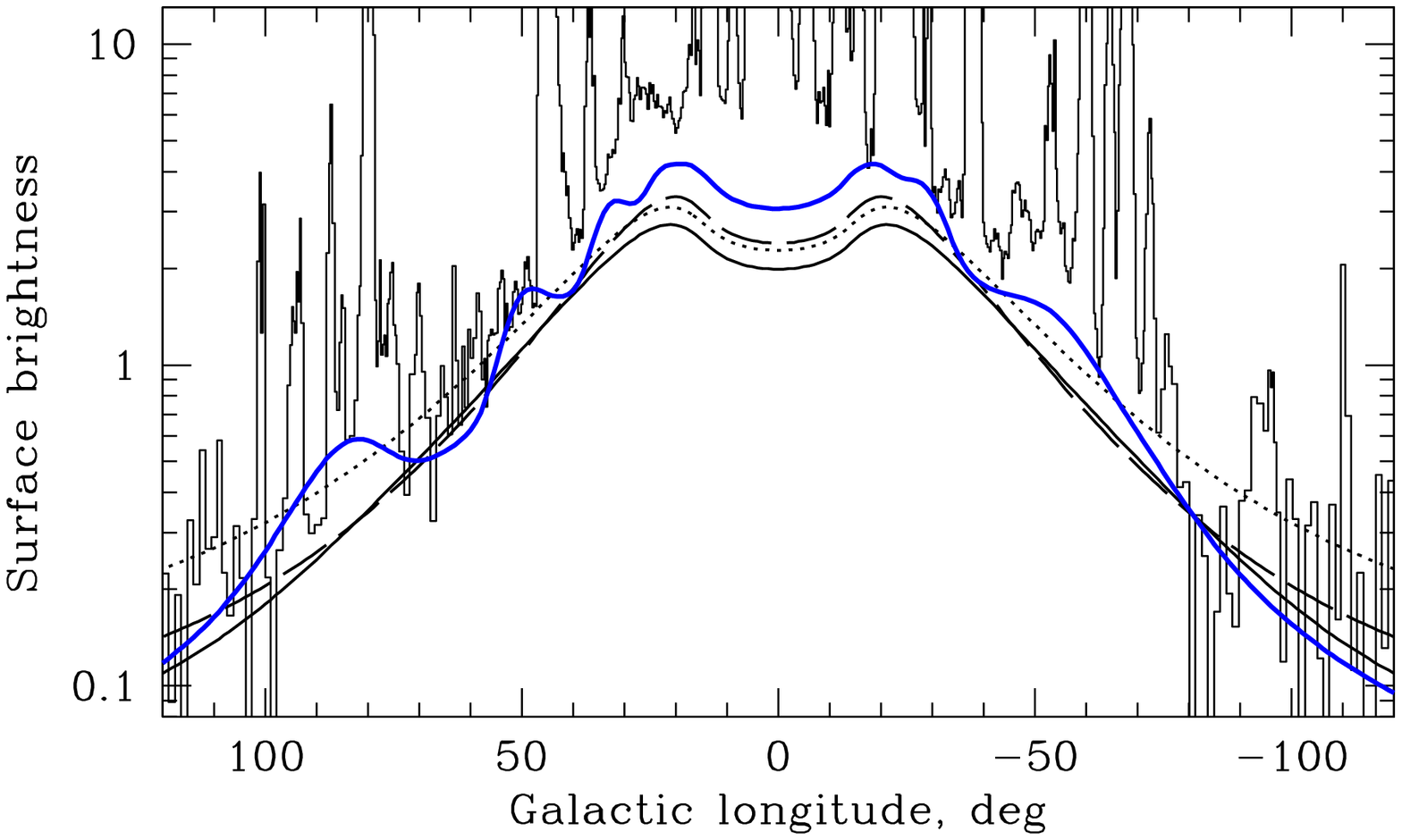}
\includegraphics[width=\columnwidth,bb=50 180 570 500,clip]{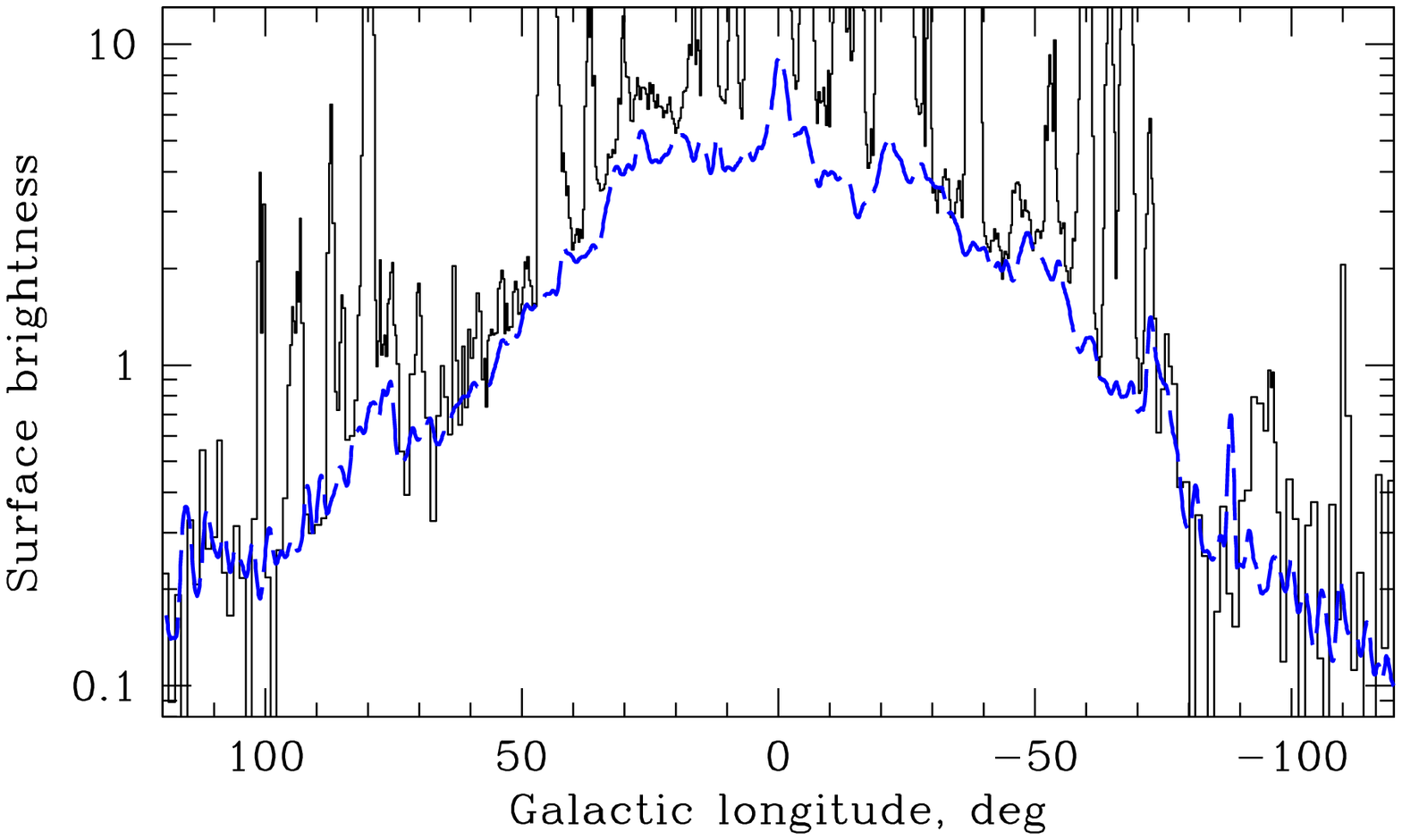}
\caption{{\sl Upper panel:} X-ray intensity profile (in units of
$10^{-11}$ \flux\ deg$^{-2}$) along the Galactic plane ($|b|<0.5$).
The dashed line shows a model of infinite exponential disk with
scale length $R_{\rm disk}=2.0$ kpc, the dotted line is a model with
$R_{\rm disk}=2.5$ kpc, solid line is a model with $R_{\rm
disk}=2.5$ kpc with disk truncation at $R_{\rm max}=10$ kpc. The
thick solid line shows an intensity profile corresponding to a
simple spiral structure of the Galaxy consisting of four logarithmic
spirals with a pitch angle 15$^\circ$. The intensity peaks manifest
the tangents to the spiral arms. {\sl Lower panel:} The solid line
is again the X-ray surface brightness profile of the Galaxy
($|b|<0.5$), the dashed line is the near-infrared surface brightness
distribution measured by COBE/DIRBE (3.5 $\mu$m) corrected for
interstellar extinction and multiplied by a factor $0.26
\times10^{-11}$ \flux deg$^{-2}$ (MJy/sr)$^{-1}$ (see text). The
near-infrared map was convolved with the response of the RXTE/PCA
collimator ($\sim1^\circ$)  } \label{slice_plane}
\end{figure}

Following \cite{bahcall80,kent91,freudenreich96,dehnen98}, we can
parameterize the volume emissivity in the disk $\rho_{\rm disk}$
by two exponentials (of Galactocentric distance $R$ and height
above the disk plane $z$) with a central depression (parameter
$R_{\rm m}$). At radii $R<R_{\rm max}$ the volume emissivity is given by

\begin{equation}
\rho_{\rm disk}=\rho_{0, \rm disk} \,\, \exp\left[-\left({R_{\rm
m}\over{R}}\right)^3-{R\over{R_{\rm disk}}} - {z\over{z_{\rm
disk}}}\right],
\label{disk_model}
\end{equation}
while $\rho_{\rm disk}=0$ at $R>R_{\rm max}$.

The very limited information that we have about the emissivity of the
disk near the Galactic Center (due to bright source contamination)
does now allow us to determine the parameter $R_{\rm m}$. We
hence fixed it at the value $R_{\rm m}=3.0$ kpc, which is
approximately the size of the central stellar disk depression
supposedly induced by the bulge/bar of approximately
the same size \citep[e.g.][]{freudenreich96,freudenreich98}. We note
that a central disk depression within approximately $R=3.0$~kpc is also
evident on the Galactic plane map of 6.7~keV line emission obtained by
\cite{yamauchi93}, which is much less affected by bright point sources
than the GRXE continuum map obtained by RXTE.

In Fig.~\ref{slice_plane} we show a number of modelled disk
intensity profiles: a) an infinite disk with an exponential scale
length $R_{\rm disk}=2.5$ kpc ($R_{\rm max}=\infty$, dotted line),
b) a disk with $R_{\rm disk}=2.5$ kpc truncated at $R_{\rm max}=10$
kpc (solid line) and c) an infinite disk with $R_{\rm disk}=2.0$ kpc
(dashed line). Comparison with the measured surface brightness
profile demonstrates that the first model provides the worst fit: it
predicts too much emission at large distances from the Galactic
Center. Both the infinite disk with $R_{\rm disk}=2.0$ kpc and the
finite disk with $R_{\rm disk}=2.5$ kpc and $R_{\rm max}=10$ kpc
match the observed profile much better. In the subsequent analysis
we will assume that the GRXE disk has a scale length of $R_{\rm
disk}=2.5$ kpc and is truncated at $R_{\rm max}=10$ kpc. These
values are very close to the parameters of the Galactic
stellar disk \citep[e.g.][]{freudenreich96,binney97,freudenreich98}.

The GRXE is not detectable with RXTE/PCA at longitudes
$|l|>80^\circ$, with an upper limit $\sim 3\times 10^{-12}$ \flux\
deg$^{-2}$ for any line of sight. We point out that measurement of
the (weak) ridge emission at these longitudes by an instrument with
poor angular resolution can be strongly affected by the presence of
relatively bright discrete sources. For example, the published
CGRO/OSSE measurement of GRXE at $l=95^\circ$ \citep{skibo97} is
possibly dominated by emission from the active galactic nucleus IGR
J21247+5058 recently discovered by INTEGRAL \citep{masetti04}.

\subsubsection{Possible imprint of the disk spiral structure}

There are indications of the GRXE being asymmetric with respect to the Galactic
Center. In particular the ridge emission appears stronger at
$l=-60^\circ$ than at $l=60^\circ$ (see Fig.~\ref{map} and
Fig.~\ref{slice_plane}). This asymmetry is probably caused
by a significant enhancement of the GRXE in spiral arms with respect to
inter-arm regions. Such arm-interarm contrast of the volume
emissivity is seen on near-infrared maps of stellar disks in
spiral galaxies \cite[e.g.][]{gonzalez96,drimmel01}. The observed
enhancement of the GRXE in the direction of $l=-60^\circ$ could then
be a signature of the tangent to the Crux spiral arm. In
Fig.~\ref{slice_plane} we show an example of GRXE longitudinal
intensity profile corresponding to a Galactic disk with spiral
structure (thick solid line). For this model we assumed a simple
four-logarithmic-arms spiral structure \cite[e.g.][]{vallee95} with a
pitch angle of $15^\circ$. The width of the spiral arms was assumed to be
600 pc, and the arm-interarm volume emissivity contrast was assumed to
be 2.0.

Summarizing all of the above we can conclude that the intensity
of the disk component of the GRXE follows the near-infrared brightness
distribution of the Galaxy. To strengthen this conclusion
we present in Fig.~\ref{slice_plane}b the near-infrared brightness profile
of the Galaxy at $|b|<0.5^\circ$ obtained by COBE/DIRBE
(zodi-subtracted mission average map provided by the LAMBDA archive of
the Goddard Space Flight Center, http://lambda.gsfc.nasa.gov). We have
chosen the spectral band centered on 3.5$\mu$m, for which the interstellar
extinction is fairly small, and made the simplest correction for the extinction
using the map of interstellar HI gas of \cite{dickey90} and
extinction law by \cite{rieke85}. The infrared profile was
additionally convolved with the PCA collimator response. A very good
correspondence between  the GRXE and near-infrared intensity profiles
is apparent.

\subsubsection{Vertical extent}

\begin{figure}
\includegraphics[width=\columnwidth,bb=40 180 570 700,clip]{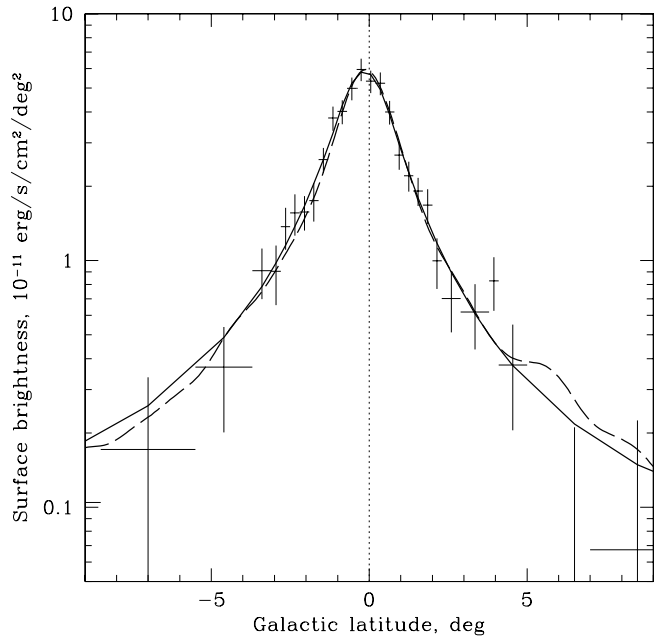}
\caption{GRXE intensity profile in the $\sim 1^\circ$-wide slide
around $l=20.2^\circ$ perpendicular to the Galactic plane. The solid
line shows the best-fit model of the GXRE disk component (see text).
The dashed line shows the near-infrared (3.5 $\mu$m, extinction
corrected) brightness distribution in units of MJy/sr multiplied by
a factor 0.26. } \label{slice_scutum}
\end{figure}

As was already mentioned, our study of the GRXE in the Galactic plane
strongly suffers from the high density of bright X-ray sources and
limited angular resolution of RXTE/PCA. Filtering out discrete sources in
the Galactic plane greatly reduces the amount of data at
$|b|<1^\circ$. Only a few regions are suitable for studying the
intensity profile of the GRXE perpendicular to the Galactic plane,
with evidently the best place being the region around
$l\sim20^\circ$. At this longitude there are no sources brighter than
$\sim 10^{-12}$ \flux\ \citep{sugizaki01} at $b\sim0^\circ$
and  no sources brighter than $\sim 10^{-11}$ \flux\ away from
the plane \citep{molkov04}. We present the vertical profile of the GRXE
intensity at $l=20.2^\circ$ (within a stripe of width $\sim 1^\circ$)
in Fig.~\ref{slice_scutum}.

One of the most interesting features of the vertical profile is
the displacement of its peak from $b=0^\circ$. The best-fit
peak position is $b_{\rm peak}=-0.15^\circ\pm0.02^\circ$
($1\sigma$ statistical uncertainty). Using available scans across
known persistent sources, including supernova  remnants, we estimated that the
systematic uncertainty in measuring $b_{\rm peak}$ is smaller
than $\sim 0.03^\circ$. We therefore conclude that the observed
displacement is real. A similar shift of the brightness peak was
previously observed in a near-infrared spectral band
\citep[e.g.][]{djorgovski89,freudenreich94,porcel97}, and it was
suggested to arise from the elevation of the Sun above the
disk plane.

By fitting to the measured intensity profile at $l=20.2^\circ$ the
disk model given by equation~(\ref{disk_model}), with the
Sun elevation above the Galactic plane as an additional
parameter ($z_{\rm Sun}$), we obtain the best fit shown in
Fig.~\ref{slice_scutum} by a solid line. The best-fit parameters are
presented in Table~\ref{pars}.

The measured vertical profile of the X-ray disk volume emissivity is very
similar to that of the stellar disk \cite[see e.g.][]{binney97}. To better
illustrate this we show in Fig.~\ref{slice_scutum} a vertical
profile of the near-infrared brightness of the Galactic disk measured at the
same Galactic longitude at wavelength 3.5~$\mu$m (extinction
corrected). The COBE/DIRBE infrared map was convolved with the
PCA collimator response. The agreement with the X-ray intensity profile is
excellent.

\section{Summary of the GRXE map}

Below we summarize the findings of the previous two sections.

\begin{enumerate}
\item  The GRXE consists of two major components -- a disk and a bulge.

\item The bulge/bar component of the GRXE can be well described by a triaxial
ellipsoid similar to that used for description of the Galactic
stellar bar \citep{dwek95}. The best-fit parameters of the GRXE
bulge/bar are fully compatible with those of the stellar bar.

\item The vertical intensity profile of the disk component of the GRXE
at $l=20^\circ$ exhibits a significant offset with respect to
$b=0^\circ$. This offset is compatible with that
observed for the Galactic near-infrared (e.g. 3.5$\mu$m) brightness
distribution and can be explained by the elevation of the Sun above the
Galactic plane by $\sim 20$ pc.

\item The intensity profile of the disk component of the GRXE at $l=20^\circ$
within $|b|<4^\circ$ can be described by an exponent with a scale
height $z_{\rm disk}=0.13\pm0.02$ kpc, which is compatible with the
near-infrared brightness (i.e stellar mass) distribution of the disk
\citep[e.g.][]{binney97,dehnen98}. Further away from the Galactic
plane where stellar near-infrared emission is still visible
\citep[e.g.][]{freudenreich96}, the RXTE/PCA sensitivity is not
sufficient to detect the ridge emission.

\item The longitudinal profile of the GRXE intensity is compatible
with the profile of near-infrared brightness of the Galaxy. It can be
approximated by an infinite exponential disk with a scale
length $R_{\rm disk}=2.0$ kpc or by a finite disk with a scale length $R_{\rm
disk}=2.5$ kpc and an outer boundary at $R_{\rm max}\sim 10$ kpc.
The inner stellar disk radius $R_{\rm m}\sim 3.0$ kpc is compatible with
our data but cannot be constrained by these data.

\item There are indications of asymmetry of the GRXE
distribution in the Galactic plane with respect to $l=0^\circ$. We propose
that this asymmetry might be caused by the Galactic spiral
structure. If this is the case, we expect an arm-interarm contrast of
the GRXE disk volume emissivity of $\sim$2--3, as for the
near-infrared surface brightness of stellar disks in spiral
galaxies. The current observations do not allow us to place tight
constraints on the suggested spiral structure of the GRXE.

\item The integral luminosities of the bulge and disk in the 3--20 keV
energy band are $(3.9\pm 0.5)\times 10^{37}$ erg s$^{-1}$ and
$\sim 10^{38}$ erg s$^{-1}$, respectively. The disk-to-bulge
luminosity ratio $L_{\rm x,disk}/L_{\rm x,bulge}\sim 2.5$. This
value agrees with the disk-to-bulge stellar mass ratio,
indicating that the GRXE emissivitiy per unit mass is similar in the disk and
in the bulge. We note however that a precise determination of the
luminosities of the disk and bulge components as well as of their ratio
would require an accurate modelling of the disk component across the Galaxy
and of the bulge component near the Galactic Center, and such detailed
modelling is not possible with RXTE/PCA data.
\end{enumerate}

{\sl Based on all these observational facts we can conclude
that the GRXE surface brightness (at 3--20 keV) closely follows
the near-infrared brightness of the Galaxy and thus the Galactic
stellar mass distribution.}

To calculate the ratio of X-ray luminosity to stellar mass it is
preferable to use the bulge/bar component of the GRXE, since it is
explored with RXTE/PCA in more detail compared to the disk. Assuming
a bulge/bar stellar mass of 1--1.3$\times10^{10} M_\odot$
\citep{dwek95} we find that the ratio of X-ray luminosity to stellar
mass is $L_{\rm x}/M_{\odot}\sim (3.5\pm0.5) \times 10^{27}$ erg
s$^{-1}$. The ratio of X-ray (3--20 keV) to near-infrared luminosity
at 3.5 $\mu$m (in the COBE/DIRBE band with a width $\sim$1~$\mu$m)
is $L_{\rm 3-20 keV}/L_{3-4 \mu m}=(4.1\pm0.5)\times 10^{-5}$. The
latter estimate is based on the near-infrared luminosity of the
bulge/bar  $L_{\rm 3-4 \mu m}=2.5\times10^{8} L_{\odot}$
\citep{dwek95}.

\begin{figure}
\includegraphics[height=\columnwidth,angle=-90]{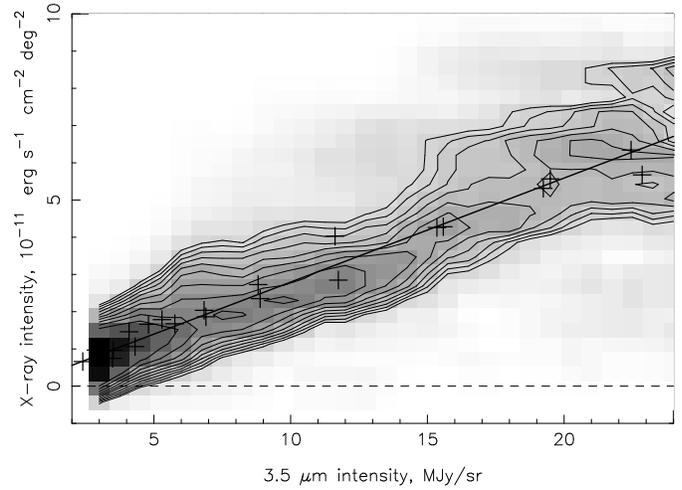}
\caption{Gray scale map and contours of isodensity of measurements
of X-ray (3--20 keV, RXTE/PCA) and near-infrared (3.5~$\mu$m,
COBE/DIRBE) intensity in multiple sky regions. Discrete X-ray
sources with flux $>1.5-2 \times 10^{-11}$ \flux\ were masked. The
solid line shows the linear correlation between near-infrared and
X-ray intensities $I_{\rm 3-20 keV}$($10^{-11}$ \flux\
deg$^{-2}$)=$0.26 \times I_{\rm 3.5\mu m}$(MJy/sr), obtained from a
vertical slice of the GRXE disk near $l=20.2^\circ$ (see
Fig.~\ref{slice_scutum}). The near-infrared and X-ray intensities
for this slice are shown by crosses.} \label{density}
\end{figure}

\begin{figure*}
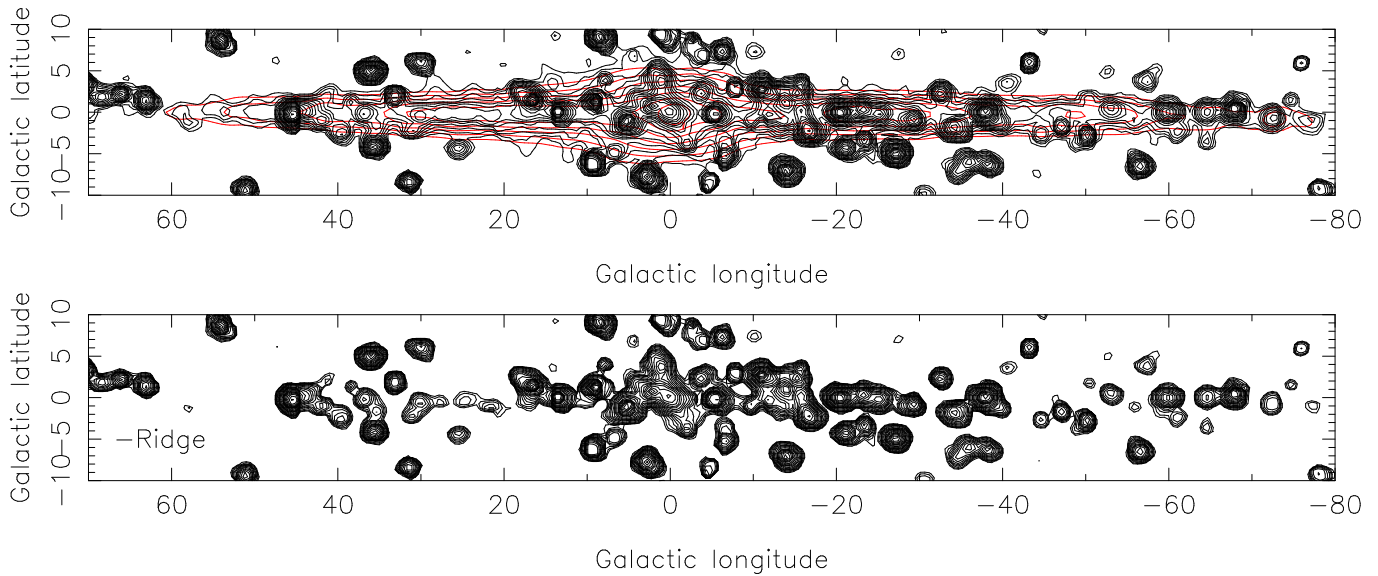

\includegraphics[height=\textwidth,angle=-90]{./ridge_contours_cobe_contours.ps}
\includegraphics[height=\textwidth,angle=-90]{./noridge_contours.ps}
\caption{{\sl Top panel:} RXTE/PCA map of the sky around the
Galactic plane in the 3--20 keV band. The red contours show the
COBE/DIRBE near-infrared map (3.5~$\mu$m) rescaled by a factor
$I_{\rm 3-20 keV}$($10^{-11}$ \flux\ deg$^{-2}$)/$I_{\rm 3.5\mu
m}$(MJy/sr)=$0.26\pm0.05$ and convolved with the PCA collimator
response. The contour levels are logarithmically spaced with a
factor of 1.4, with the lowest contour corresponding to an X-ray
intensity of $10^{-11}$ \flux\ deg$^{-2}$. {\sl Lower panel:} the
same RXTE/PCA map with subtracted rescaled near-infrared map.
Contour levels are the same as for the upper panel.} \label{map_noridge}
\end{figure*}

To illustrate the nearly perfect correlation between the GRXE and
the Galactic near-infrared brightness we show in Fig.~\ref{density}
a scatter plot of X-ray surface brightness vs. near-infrared
brightness, constructed by dividing the sky into many small regions.
To obtain this diagram, detected point X-ray sources were masked and
the infrared sky map, having much better angular resolution than the
X-ray map, was convolved with the response of the RXTE/PCA
collimator. The scatter plot is shown as a gray scale map of density
of measurements with superposed isodensity contours. The density is
calculated within boxes of size (3MJy/sr)$\times (1.5\times
10^{-11}$ \flux\ deg$^{-2}$). The observed stripe of enhanced
density signifies the correlation of near-infrared and GRXE
intensities. The vertical width of the stripe can be fully accounted
for by the uncertainty of X-ray measurements.

The one-to-one correspondence between the GXRE map and the Galactic
near-infrared brightness map permits to express GRXE intensity through
infrared intensity at wavelength 3.5$\mu$m for any line of sight:
$I_{\rm 3-20 keV}$($10^{-11}$ \flux\ deg$^{-2}$)=$(0.26\pm0.05)
\times I_{\rm 3.5\mu m}$(MJy/sr). This linear relation is derived by
fitting the vertical slice of the GRXE at $l=20^\circ$ to the
corresponding profile of near-infrared surface brightness (see
Fig.~\ref{slice_scutum}) and is shown in Fig.~\ref{density} (solid
line) together with individual measurements for the
$l=20^\circ$ slice (crosses).

The above discussion suggests that the extinction corrected
COBE/DIRBE near-infrared (3.5$ \mu$m)  map of the sky should be
nearly identical to the GRXE map upon scaling by a factor $I_{\rm
3-20 keV}$($10^{-11}$ \flux\ deg$^{-2}$)/$I_{\rm 3.5\mu
m}$(MJy/sr)=0.26. Fig.~\ref{map_noridge} demonstrates that this is
indeed the case. Subtracting the rescaled near-infrared map from the
observed X-ray brightness map of the Galaxy (Fig. \ref{map_noridge},
upper panel) leaves only point-like X-ray sources
(Fig.~\ref{map_noridge}, lower panel).

\section{Broad-band spectrum of the GRXE}

The infrared brightness of the Galaxy sharply rises within
10\arcmin\ of the Galactic Center (Sgr A$^*$) because of the
so-called nuclear stellar cluster
\cite[e.g.][]{genzel87,launhardt02}. We showed above that
regardless of the origin of the GRXE its surface brightness traces the
near-infared surface brightness. One may therefore anticipate a
sharp rise of the GRXE in the inner 10\arcmin\ of the Galaxy. Such an
X-ray intensity spike has indeed been observed with Chandra
\citep{muno03} and it was demonstrated that the total X-ray flux from
the inner 10\arcmin\ of the Galaxy is not dominated by bright point sources
\cite[e.g.][]{muno03,neronov05}.

If observed by a hard X-ray telescope with moderate angular
resolution, such as IBIS aboard INTEGRAL (angular resolution
12\arcmin, \citealt{winkler03}), the GRXE central cusp will be
perceived as a point-like source in the Galactic Center. The total flux
from this source, assuming that the ratio of the GRXE emissivity to the
stellar mass ($L_{\rm x}/M_{\odot}$) is the same as in the other parts
of the Galaxy, will be determined by the total mass of the nuclear
stellar cluster. The innermost 30~pc (corresponding to 12\arcmin\ at
the Galactic Center distance) of the Galaxy enclose $M_{\rm
cusp}\sim 10^{8} M_{\odot}$ of stars. This predicts a 3--20 keV
luminosity of $L_{\rm x}\sim 3.5\times 10^{27} M_{\rm cusp}\sim
4\times10^{35}$ erg s$^{-1}$.

If we further assume that the GRXE spectrum in the 3--60 keV range is
a power law with a photon index $\Gamma=2.1$ (as measured in the
3--20~keV range for the bulge component of the GRXE, \citealt{mikej03}), we can
roughly predict the hard X-ray luminosity (20--60 keV) of the GRXE
from the central cusp: $L_{\rm 20-60 keV,~estimate}\sim
2\times10^{35}$ \lum. This proves to be only a factor of 2 less than the
hard X-ray luminosity of the Galactic Center "point  source" measured
by INTEGRAL/IBIS \citep{neronov05,belanger05}, which may be considered
a good agreement in view of uncertainty concerning the mass of
the nuclear stellar cluster and the solid angle subtended by
the INTEGRAL source. This also leaves open the possibility that the
GRXE from the nuclear region of the Galaxy, which is known to be
peculiar in many respects, may be somewhat different from the rest of
the Galaxy.

\begin{figure}
\includegraphics[width=\columnwidth]{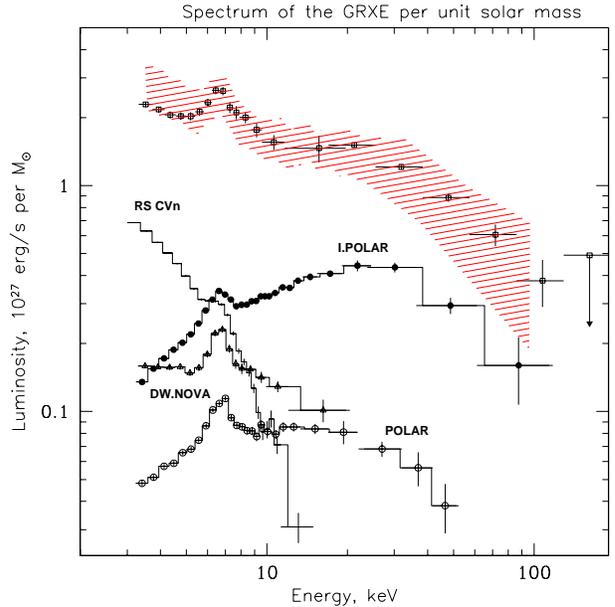}
\caption{GRXE broad-band spectrum (squares) and spectra of its main
contributors divided by 2 for clarity. The data points in the 3--20
keV band (RXTE/PCA) were converted to unit-stellar-mass emissivity
(based on the correlation of the GRXE with the near-infrared
brightness, see text). The data points in the 20--100 keV band show
the spectrum of the Galactic Center source IGR J17456$-$2901
measured by INTEGRAL/IBIS divided by the estimated total mass in
stars ($\sim 10^{8} M_\odot$) contained in the nuclear region
($\sim$30 pc around Sgr A$^*$). The INTEGRAL/IBIS spectrum was
additionally multipled by a factor $0.6$ to match the RXTE/PCA
spectrum near 20~keV. Also shown are typical spectra of X-ray source
classes expected to significantly contribute to the GRXE:
intermediate polars (V1223 Sgr, filled circles), polars (AM Her,
open circles), dwarf novae (SU UMa, triangles), and coronally active
binaries (V711 Tau). These spectra are plotted with normalizations
corresponding to their expected relative contributions to the GRXE
(derived from the local statistics of X-ray sources) divided by 2
(for better visibility). The individual source spectra were obtained
by the instruments PCA (3--20 keV) and HEXTE (20--100 keV) aboard
RXTE. The shaded region shows a sum of these spectra reflecting
uncertainties in the individual spectra and their relative weights.}
\label{spectrum_composite}
\end{figure}

Given the good correspondence between the predicted hard X-ray flux
from the central stellar cusp (based on the correlation between the
GRXE and the Galactic stellar mass) and the measured flux from the
Galactic Center hard X-ray source, it is interesting to attach the
INTEGRAL/IBIS spectrum of the Galactic Center source at energies
above 20~keV to the spectrum of the large-scale GRXE measured below
20~keV by RXTE/PCA. Both spectra need to be normalized to the same
stellar mass. We show in Fig.~\ref{spectrum_composite} the resulting
combined spectrum covering a broad energy range from 3 to $\sim
100$~keV. This spectrum may to a first approximation be regarded as
a broad-band spectrum of the GRXE. However, one should keep in mind
the possibility that the actual hard X-ray spectrum of the GRXE from
regions away from the Galactic Center may prove somewhat different,
since the Galactic Center is a peculiar region.

\section{Galactic ridge X-ray emission as a superposition of point sources}

In the previous sections we presented evidence that the GRXE volume
emissivity traces the Galactic stellar density and estimated the
ridge X-ray emissivity (3--20~keV) per unit stellar mass as $L_{\rm
x}\sim (3.5\pm0.5) \times 10^{27}$ erg s$^{-1} M_\odot^{-1}$. It is
interesting to compare this number with the cumulative X-ray
emissivity of known classes of X-ray sources.

It is known \cite[see e.g.][]{sazonov05} that there is a clear
tendency of spectral hardening with increasing source luminosity.
Specifically, the weakest known X-ray sources ($<10^{31}$ erg
s$^{-1}$) are characterized by multi-temperature thermal spectra
peaking in the soft X-ray regime \citep[stellar coronal sources, see
e.g.][]{schmitt90,gudel04}, while the brightest sources contributing
to the GRXE ($\sim 10^{33-34}$ erg s$^{-1}$) have very hard spectra
with significant energy output above 10~keV \cite[intermediate
polars, see e.g.][]{suleimanov05}. Therefore, it is important to
consider the same energy band (3--20 keV) as was used in our GRXE study.

We have recently used the RXTE slew survey of the sky at high
Galactic latitude to construct an X-ray (3--20~keV) luminosity
function of nearby ($\sim 1$~kpc) sources, covering a broad range in
luminosity from coronally active stellar binaries to white dwarf
binaries \citep{sazonov05}. Based on this luminosity function it is
straightforward to estimate the contribution of point sources to the
GRXE measured by RXTE in the same spectral band.

The total 3-20 keV emissivity of local Galactic X-ray sources per
unit stellar mass $L_{\rm x,local}/M= (5.3\pm 1.5)\times10^{27}$
\lum\ $M_\odot^{-1}$ or $(6.2\pm 1.5)\times10^{27}$ \lum\
$M_\odot^{-1}$, if the contribution of young coronal stars is
excluded or included, respectively \citep{sazonov05}. The bulk of
the local X-ray emissivity is produced by coronally active late-type
binaries and cataclysmic variables. These classes of sources
represent relatively old stellar populations and their number
density is thus expected to closely trace the overall stellar
density in the Galaxy. On the other hand, the relative fraction of
young stellar objects is expected to vary strongly from one Galactic
region to another, so their locally estimated emissivity may not
well represent the Galaxy as a whole. We find that the local X-ray
emissivity, excluding the (small) contribution of young coronal
stars, agrees within the uncertainties with the GRXE emissivity,
$(3.5\pm0.5)\times 10^{27}$ erg s$^{-1} M_\odot^{-1}$, found in this
paper. This suggests that the bulk of the GRXE may be composed of
weak X-ray sources of known classes, mostly coronally active
binaries and cataclysmic variables.

If the GRXE is indeed superposed of known populations of X-ray
sources, then its energy spectrum must be a sum of the spectra of
these sources. In Fig. \ref{spectrum_composite} we compare the
measured spectrum of the GRXE with typical spectra of those classes
of sources that are expected to contribute significantly to the
GRXE. Also a composite spectrum is shown (the dashed region), which
is a weighted sum of the individual spectra. The weights describing
the fractional contributions of different types of sources were
fixed at the values determined by \cite{sazonov05} for X-ray sources
in the Solar neighbourhood, namely intermediate polars : polars :
dwarf novae : coronally active binaries  -- 1:0.2:0.6:2.0. As can be
seen in Fig.~\ref{spectrum_composite}, the composite spectrum turns
out to be very similar to the GRXE spectrum.

\section{On the way of resolving the GRXE}

It is impossible to place strong constraints on the contribution of
truly diffuse emission to the GRXE based on the X-ray luminosity
function of local  sources. First, there remains significant
uncertainty as concerns the local Galactic X-ray volume emissivity.
But even if future dedicated survey missions like ROSITA
\cite[e.g][]{rosita} will determine this quantity more accurately,
there will still remain significant uncertainty owing to the fact
that the locally determined unit-stellar-mass X-ray emissivity may
represent other parts of the Galaxy only approximately.

It appears that the only possibility to tightly constrain the
possible contribution of truly diffuse emission to the GRXE is to
detect the weakest point sources at $\sim 10$~kpc distances and to
subtract their contribution from the GRXE. In pursuing this goal,
one will likely face the following observational challenges.

\subsection{Surface brightness limit}

Far away from the Galactic Center and the Galactic plane the GRXE is
characterized by low surface brightness. It may be anticipated that
below a certain intensity level the GRXE will be lost in the cosmic
variance or small scale fluctuations of the CXB.

This implies that it should be difficult to study the GRXE in
regions characterized by surface brightness less than $\sim
10^{-11}$ \flux\ deg$^{-2}$ (CXB intensity is  $\sim 2\times
10^{-11}$ \flux\ deg$^{-2}$). Higher GRXE surface brightness is
observed only in the Galactic plane and in the Galactic bulge/bar.
The best observational target in this respect is the Galactic Center
region, but there the confusion limit may become a problem.

\subsection{Confusion limit}

In regions of very high GRXE  surface brightness (e.g. the Galactic
Center region), the surface density of Galactic sources can be very
high and observations may be hampered by the confusion limit. For
example, observation with Chandra, with its excellent angular
resolution ($\sim 1''$), of the Galactic Center region will become
confusion limited  if the surface density of sources increases by a
factor of $\sim$40 (to $\sim 500$ sources arcmin$^{-2}$) compared to
published observations \citep{muno03}.

\subsection{Sensitivity limit}

The observational sensitivity should be aimed to enable the
detection of sources contributing more than e.g. 80--90\% of the
total X-ray luminosity of discrete Galactic sources (excluding bright
X-ray binaries). Using the cumulative
luminosity function of \cite{sazonov05} we can estimate that sources
with luminosities down to $L_{\rm x}\sim 10^{30}$ \lum\ will need to
be detected. For the Galactic Center region, where most sources
dominating the total flux are located
at the Galactic Center distance ($\sim8.5$ kpc), it will be necessary
to achieve a sensitivity limit of $\sim10^{-16}$ \flux, which is a factor
of $\sim 13$ improvement over the current limit \citep{muno03}.

\begin{figure}
\includegraphics[width=\columnwidth,bb=25 155 600 700,clip]{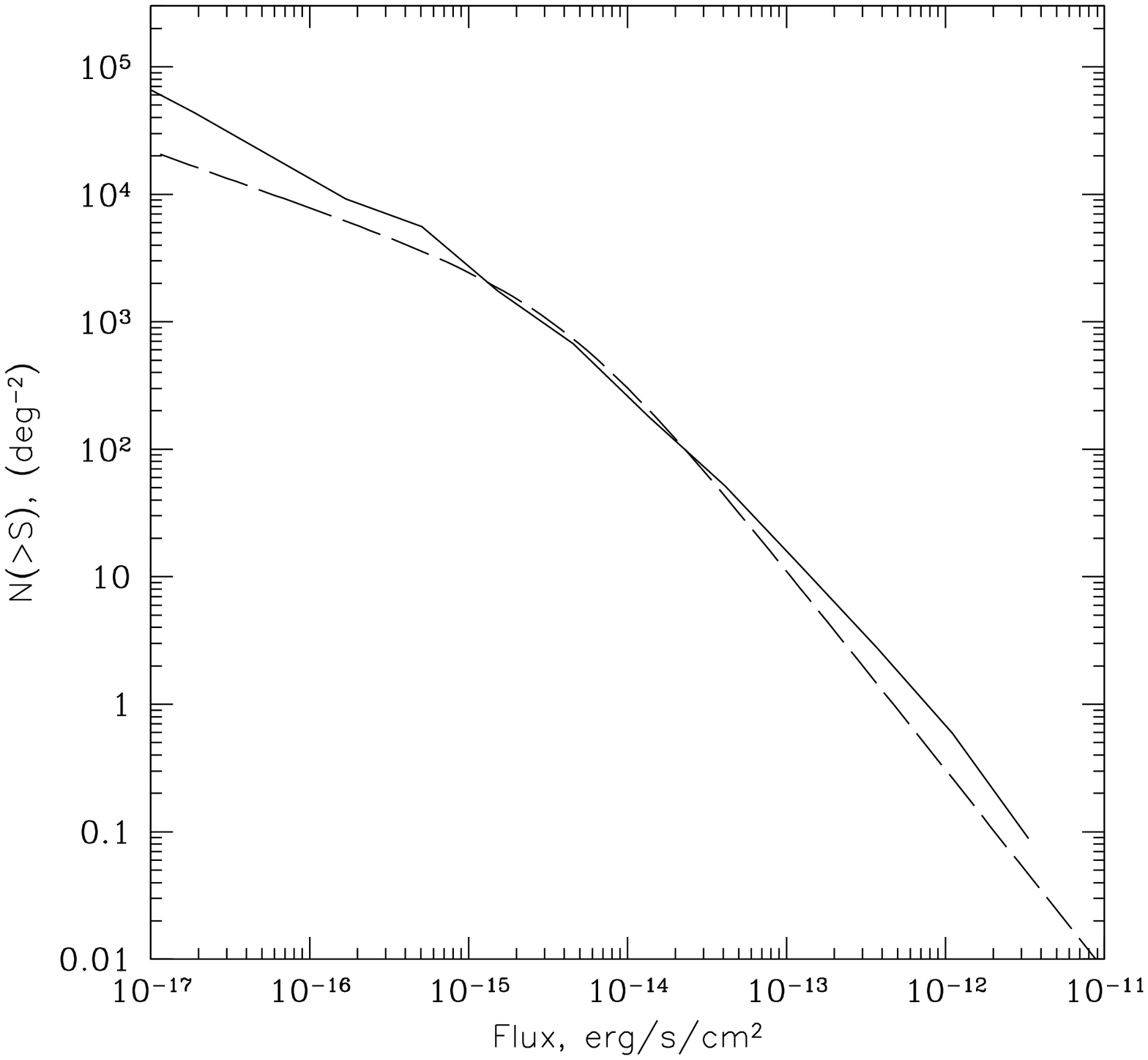}
\includegraphics[width=\columnwidth,bb=25 155 600 700,clip]{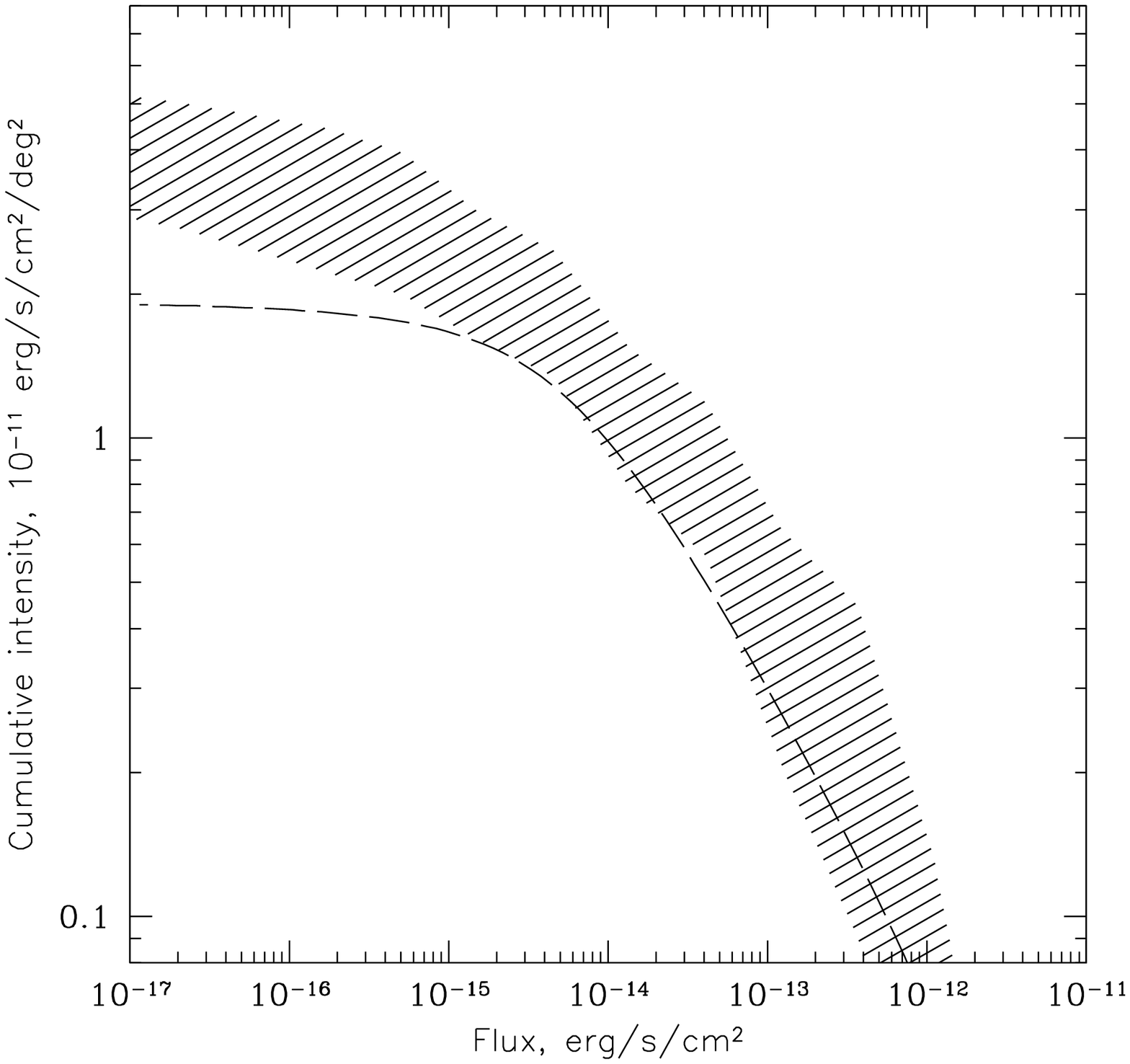}
\caption{{\sl Upper panel:} Predicted number-flux function of
Galactic (solid line) and extragalactic (dashed line) sources in the
2--10~keV energy band in the direction $l=20$, $b=0.0$. The
number-flux function of extragalactic sources is adopted from
\cite{moretti03}. {\sl Lower panel:} Cumulative surface brightness
distribution for the same populations of sources. Shaded region
reflects uncertainties in the Galactic X-ray volume emissivity}
\label{logn_logs}
\end{figure}

We can also assess the required  sensitivity for a given Galactic
plane region using our model of the Galaxy (section 3.2) and the
luminosity function of Galactic X-ray sources \citep{sazonov05}. We
use the Galactic plane region at $l=20^\circ$ as an example. The
resulting cumulative number-flux function and cumulative X-ray
surface brightness as a function of flux are presented in
Fig.~\ref{logn_logs}. These dependencies were calculated for the
standard 2--10~keV band to faciliate comparison with X-ray missions
such as Chandra and XMM-Newton. Note that the relative contributions
of different classes of sources are sensitive to the choice of
energy band. In particular, the contribution of coronally active
binaries to the GRXE at 2--10~keV is expected to be more important
compared to the 3--20 keV band (see e.g.
Fig.~\ref{spectrum_composite}). Also it should be noted that one
should be careful while comparing the presented figure with CHANDRA
results, because the effectively sensitive energy band of CHANDRA
does not continue to energies higher than $\sim5-6$ keV.

It follows from Fig.~\ref{logn_logs} that in order to resolve most
of the GRXE into discrete sources and place tight constraints on the
contribution of truly diffuse emission to the GRXE it is necessary
to reach point source flux limits $\sim 10^{-16-16.5}$ \flux.
Interestingly, such sensitivity is not unfeasible for Chandra
(XMM-Newton will be strongly limited by confusion already at fluxes
$\sim10^{-15}$ \flux). Similar sensitivities were already achieved
in deep surveys of extragalactic fields
\cite[e.g.][]{giacconi02,moretti03}. Moreover, the predicted number
density of sources above the required flux limit ($\sim 5\times
10^{4}$ deg$^{-2}$, Fig.~\ref{logn_logs}, upper panel), will not
create a confusion problem for Chandra.

At higher energies ($>20$ keV) lower sensitivities are needed,
$\sim10^{-14-13.5}$ \flux. This follows from the fact that according
to the picture presented here the dominant contribution to the GRXE
at these energies likely comes from polars and intermediate polars
with high luminosities ($L_{\rm x}\sim10^{32-34}$ \lum). The surface
density of such sources will be $10^{3-4}$ deg$^{-2}$. Both the
required sensitivity and angular resolution should be achievable
with the projected focusing hard X-ray telescopes, e.g. NuSTAR
\cite[e.g.][]{nustar}.

\section{Conclusions}

\begin{itemize}
\item In this paper we have shown that the 3--20~keV map of the GRXE
closely follows the near-infrared (3.5 $\mu$m) brightness distribution of the
Galaxy and thus traces the Galactic stellar mass distribution.

\item The GRXE map reveals the presence of the Galactic
bulge/bar in the inner 3--4 kpc. The parameters of the bar determined
from the X-ray data are fully compatible with those obtained from near-infrared
observations.

\item We demonstrated that the point-like source IGR J17456$-$2901 in the
Galactic Center observed by INTEGRAL/IBIS is likely concentrated GRXE
from the innermost $\sim 10'$. We used this fact to extend the GRXE
energy spectrum up to 100~keV.

\item Comparison of the GRXE luminosity per unit stellar mass with the
cumulative emissivity of X-ray sources in the Solar neighborhood
suggests that the bulk of the GRXE is likely superposed of
emission from weak Galactic X-ray sources, mostly cataclysmic
variables and coronally active binaries.

\item The GRXE energy spectrum in the 3--100 keV range
can be explained as a composition of spectra of different
X-ray source classes weighted in accordance with their relative
contributions to the local X-ray emissivity.

\item Based on the model of Galactic stellar mass distribution we
predict that in order to resolve $\sim 90$\% of the GRXE in the 2--10
keV band into discrete sources, it will be necessary to achieve a flux
limit $\sim10^{-16}$ \flux. This sensitivity is within Chandra
capabilities.

\item For the hard X-ray regime (20--100 keV) the requirement to
sensitivity is much less strict, since the dominant contributors to
the GRXE in this energy band are intermediate polars and polars with
relatively high X-ray luminosities. The required flux limit
$\sim10^{-14-13.5}$ \flux\ can be achieved by next-generation hard
X-ray telescopes.

\end{itemize}

\begin{acknowledgements}
Research has made use of data obtained from High Energy Astrophysics
Science Archive Research Center Online Service, provided by the
NASA/Goddard Space Flight Center. We acknowledge the use of the
Legacy Archive for Microwave Background Data Analysis (LAMBDA).
Support for LAMBDA is provided by the NASA Office of Space Science.
MR thank Ralf Launhardt for his advices during the work with
COBE/DIRBE data and Andrii Neronov for provided information about
XMM-Newton results on the Galactic Center.
\end{acknowledgements}


\begin{thebibliography}{}

\bibitem[\protect\citeauthoryear{Bahcall \&
Soneira}{1980}]{bahcall80} Bahcall J.~N., Soneira R.~M., 1980,
ApJS, 44, 73

\bibitem[\protect\citeauthoryear{Belanger et
al.}{2005}]{belanger05} Belanger G., et al., 2005, astro,
arXiv:astro-ph/0508128

\bibitem[\protect\citeauthoryear{Binney, Gerhard, \&
Spergel}{1997}]{binney97} Binney J., Gerhard O., Spergel D.,
1997, MNRAS, 288, 365

\bibitem[\protect\citeauthoryear{Bleach et al.}{1972}]{bleach72}
Bleach R.~D., Boldt E.~A., Holt S.~S., Schwartz D.~A., Serlemitsos P.~J.,
1972, ApJ, 174, L101


\bibitem[\protect\citeauthoryear{Blitz \&
Spergel}{1991}]{blitz91} Blitz L., Spergel D.~N., 1991, ApJ,
379, 631

\bibitem[\protect\citeauthoryear{Bradt et al.}{1993}]{rxte} Bradt H., Rotshild R. \&
Swank J. 1993, Astron. Astrophys. Suppl. Ser.  97, 355

\bibitem[\protect\citeauthoryear{Cooke, Griffiths, \&
Pounds}{1970}]{cooke70} Cooke B.~A., Griffiths R.~E., Pounds
K.~A., 1970, IAUS, 37, 280

\bibitem[\protect\citeauthoryear{Dehnen \&
Binney}{1998}]{dehnen98} Dehnen W., Binney J., 1998, MNRAS, 294,
429

\bibitem[\protect\citeauthoryear{Djorgovski \&
Sosin}{1989}]{djorgovski89} Djorgovski S., Sosin C., 1989, ApJ, 341,
L13

\bibitem[\protect\citeauthoryear{Dickey \&
Lockman}{1990}]{dickey90} Dickey J.~M., Lockman F.~J., 1990,
ARA\&A, 28, 215

\bibitem[\protect\citeauthoryear{Drimmel \&
Spergel}{2001}]{drimmel01} Drimmel R., Spergel D.~N., 2001, ApJ,
556, 181

\bibitem[\protect\citeauthoryear{Dwek et al.}{1995}]{dwek95}
Dwek E., et al., 1995, ApJ, 445, 716

\bibitem[\protect\citeauthoryear{Ebisawa et
al.}{2001}]{ebisawa01} Ebisawa K., Maeda Y., Kaneda H., Yamauchi
S., 2001, Sci, 293, 1633

\bibitem[\protect\citeauthoryear{Ebisawa et
al.}{2005}]{ebisawa05} Ebisawa K., et al., 2005, astro,
arXiv:astro-ph/0507185

\bibitem[\protect\citeauthoryear{Freudenreich et
al.}{1994}]{freudenreich94} Freudenreich H.~T., et al., 1994, ApJ,
429, L69

\bibitem[\protect\citeauthoryear{Freudenreich}{1996}]{freudenreich96}
Freudenreich H.~T., 1996, ApJ, 468, 663


\bibitem[\protect\citeauthoryear{Freudenreich}{1998}]{freudenreich98}
Freudenreich H.~T., 1998, ApJ, 492, 495

\bibitem[\protect\citeauthoryear{Genzel \&
Townes}{1987}]{genzel87} Genzel R., Townes C.~H., 1987, ARA\&A, 25,
377

\bibitem[\protect\citeauthoryear{Giacconi et al.}{1962}]{giacconi62}Giacconi R., Gursky H., Paolini R., Rossi B. 1962, Phys.Rev.Lett. 9, 439

\bibitem[\protect\citeauthoryear{Giacconi et al.}{2002}]{giacconi02} Giacconi R.,  Zirm A., Wang J. et al. 2002, ApJS, 139, 369

\bibitem[\protect\citeauthoryear{Gonzalez \&
Graham}{1996}]{gonzalez96} Gonzalez R.~A., Graham J.~R., 1996,
ApJ, 460, 651


\bibitem[\protect\citeauthoryear{Grimm, Gilfanov, \&
Sunyaev}{2002}]{grimm02} Grimm H.-J., Gilfanov M., Sunyaev R.,
2002, A\&A, 391, 923

\bibitem[\protect\citeauthoryear{G{\" u}del}{2004}]{gudel04}
G{\" u}del M., 2004, A\&ARv, 12, 71


\bibitem[\protect\citeauthoryear{Hands et al.}{2004}]{hands04}
Hands A.~D.~P., Warwick R.~S., Watson M.~G., Helfand D.~J., 2004, MNRAS,
351, 31

\bibitem[\protect\citeauthoryear{Harrison et al.}{2004}]{nustar} Harrison F., NuSTAR Science Team, 2004, HEAD,
8,4105


\bibitem[\protect\citeauthoryear{Iwan et al.}{1982}]{iwan82}
Iwan D., Shafer R.~A., Marshall F.~E., Boldt E.~A., Mushotzky R.~F.,
Stottlemyer A., 1982, ApJ, 260, 111

\bibitem[\protect\citeauthoryear{Kent, Dame, \&
Fazio}{1991}]{kent91} Kent S.~M., Dame T.~M., Fazio G., 1991,
ApJ, 378, 131

\bibitem[\protect\citeauthoryear{Koyama et al.}{1986}]{koyama86}
Koyama K., Makishima K., Tanaka Y., Tsunemi H., 1986, PASJ, 38, 121

\bibitem[\protect\citeauthoryear{Koyama et al.}{1989}]{koyama89}
Koyama K., Awaki H., Kunieda H., Takano S., Tawara Y., 1989, Natur,
339, 603

\bibitem[\protect\citeauthoryear{Launhardt, Zylka, \&
Mezger}{2002}]{launhardt02} Launhardt R., Zylka R., Mezger P.~G.,
2002, A\&A, 384, 112


\bibitem[\protect\citeauthoryear{Lebrun et al.}{2004}]{lebrun04}
Lebrun F., et al., 2004, Natur, 428, 293

\bibitem[\protect\citeauthoryear{Lindqvist, Habing, \&
Winnberg}{1992}]{lindqvist92} Lindqvist M., Habing H.~J., Winnberg
A., 1992, A\&A, 259, 118

\bibitem[\protect\citeauthoryear{Masetti et
al.}{2004}]{masetti04} Masetti N., Palazzi E., Bassani L., Malizia
A., Stephen J.~B., 2004, A\&A, 426, L41

\bibitem[\protect\citeauthoryear{Moretti et
al.}{2003}]{moretti03} Moretti A., Campana S., Lazzati D.,
Tagliaferri G., 2003, ApJ, 588, 696


\bibitem[\protect\citeauthoryear{Markwardt et al.}{2002}]{craig02} Markwardt, C., Jahoda, K. \&  Smith, D.
 A. 2002,
http://lheawww.gsfc.nasa.gov/users/craigm/pca-bkg/bkg-users.html

\bibitem[\protect\citeauthoryear{Molkov et al.}{2004}]{molkov04}
Molkov S.~V., Cherepashchuk A.~M., Lutovinov A.~A., Revnivtsev M.~G.,
Postnov K.~A., Sunyaev R.~A., 2004, AstL, 30, 534


\bibitem[\protect\citeauthoryear{Mukai \&
Shiokawa}{1993}]{mukai93} Mukai K., Shiokawa K., 1993, ApJ,
418, 863

\bibitem[Muno et al.(2003)]{muno03} Muno, M.~P., et al.\ 2003,
\apj, 589, 225

\bibitem[Muno et al.(2004)]{muno04} Muno, M.~P., et al.\ 2004,
\apj, 613, 326

\bibitem[\protect\citeauthoryear{Neronov et
al.}{2005}]{neronov05} Neronov A., Chernyakova M., Courvoisier
T.~J.~-., Walter R., 2005, astro, arXiv:astro-ph/0506437


\bibitem[\protect\citeauthoryear{Ottmann \&
Schmitt}{1992}]{ottmann92} Ottmann R., Schmitt J.~H.~M.~M., 1992,
A\&A, 256, 421

\bibitem[\protect\citeauthoryear{Porcel, Battaner, \&
Jimenez-Vicente}{1997}]{porcel97} Porcel C., Battaner E.,
Jimenez-Vicente J., 1997, A\&A, 322, 103

\bibitem[\protect\citeauthoryear{Predehl et al}{2003}]{rosita} Predehl, P., Friedrich,
P., Hasinger, G., Pietschon, W., \& the ROSITA Team 2003,
Astronomische Nachrichten, 324, 128


\bibitem[\protect\citeauthoryear{Revnivtsev}{2003}]{mikej03}
Revnivtsev M., 2003, A\&A, 410, 865

\bibitem[\protect\citeauthoryear{Revnivtsev et
al.}{2003}]{mikej_cxb} Revnivtsev M., Gilfanov M., Sunyaev R.,
Jahoda K., Markwardt C., 2003, A\&A, 411, 329

\bibitem[\protect\citeauthoryear{Revnivtsev et
al.}{2004}]{mikej_xss} Revnivtsev M., Sazonov S., Jahoda K.,
Gilfanov M., 2004, A\&A, 418, 927

\bibitem[\protect\citeauthoryear{Rieke \&
Lebofsky}{1985}]{rieke85} Rieke G.~H., Lebofsky M.~J., 1985,
ApJ, 288, 618

\bibitem[\protect\citeauthoryear{Sazonov \&
Revnivtsev}{2004}]{sazonov04} Sazonov S.~Y., Revnivtsev M.~G., 2004,
A\&A, 423, 469v

\bibitem[\protect\citeauthoryear{Sazonov et al.}{2005}]{sazonov05}
Sazonov S., et al., 2005, A\&A submitted, astro-ph/??

\bibitem[\protect\citeauthoryear{Schmitt et
al.}{1990}]{schmitt90} Schmitt J.~H.~M.~M., Collura A., Sciortino
S., Vaiana G.~S., Harnden F.~R., Rosner R., 1990, ApJ, 365, 704

\bibitem[\protect\citeauthoryear{Skibo et al.}{1997}]{skibo97}
Skibo J.~G., et al., 1997, ApJ, 483, L95

\bibitem[\protect\citeauthoryear{Sugizaki et
al.}{2001}]{sugizaki01} Sugizaki M., Mitsuda K., Kaneda H.,
Matsuzaki K., Yamauchi S., Koyama K., 2001, ApJS, 134, 77

\bibitem[\protect\citeauthoryear{Suleimanov, Revnivtsev, \&
Ritter}{2005}]{suleimanov05} Suleimanov V., Revnivtsev M., Ritter
H., 2005, A\&A, 435, 191

\bibitem[\protect\citeauthoryear{Sunyaev, Markevitch, \&
Pavlinsky}{1993}]{sunyaev93} Sunyaev R.~A., Markevitch M., Pavlinsky
M., 1993, ApJ, 407, 606

\bibitem[\protect\citeauthoryear{Tanaka, Miyaji, \&
Hasinger}{1999}]{tanaka99} Tanaka Y., Miyaji T., Hasinger G.,
1999, AN, 320, 181

\bibitem[\protect\citeauthoryear{Tanaka}{2002}]{tanaka02} Tanaka
Y., 2002, A\&A, 382, 1052

\bibitem[\protect\citeauthoryear{Terrier et al.}{2004}]{terrier04} Terrier R.,
Lebrun F., Belanger G. et al. 2004, Proceedings of the 5th INTEGRAL Workshop, Munich 16-20 February 2004. ESA SP-552, astro-ph/0405207

\bibitem[\protect\citeauthoryear{Vallee}{1995}]{vallee95} Vallee
J.~P., 1995, ApJ, 454, 119


\bibitem[\protect\citeauthoryear{Valinia \&
Marshall}{1998}]{valinia98} Valinia A., Marshall F.~E., 1998,
ApJ, 505, 134


\bibitem[\protect\citeauthoryear{Worrall et
al.}{1982}]{worrall82} Worrall D.~M., Marshall F.~E., Boldt
E.~A., Swank J.~H., 1982, ApJ, 255, 111

\bibitem[\protect\citeauthoryear{Worrall \&
Marshall}{1983}]{worrall83} Worrall D.~M., Marshall F.~E., 1983,
ApJ, 267, 691

\bibitem[\protect\citeauthoryear{Warwick, Pye, \&
Fabian}{1980}]{warwick80} Warwick R.~S., Pye J.~P., Fabian A.~C.,
1980, MNRAS, 190, 243

\bibitem[\protect\citeauthoryear{Warwick et
al.}{1985}]{warwick85} Warwick R.~S., Turner M.~J.~L., Watson
M.~G., Willingale R., 1985, Natur, 317, 218

\bibitem[\protect\citeauthoryear{Warwick et
al.}{1988}]{warwick88} Warwick R.~S., Norton A.~J., Turner
M.~J.~L., Watson M.~G., Willingale R., 1988, MNRAS, 232, 551

\bibitem[\protect\citeauthoryear{Weiland et
al.}{1994}]{weiland94} Weiland J.~L., et al., 1994, ApJ, 425, L81

\bibitem[\protect\citeauthoryear{Winkler et
al.}{2003}]{winkler03} Winkler C., et al., 2003, A\&A, 411, L1


\bibitem[\protect\citeauthoryear{Yamauchi et
al.}{1990}]{yamauchi90} Yamauchi S., Kawada M., Koyama K., Kunieda
H., Tawara Y., 1990, ApJ, 365, 532

\bibitem[\protect\citeauthoryear{Yamauchi \&
Koyama}{1993}]{yamauchi93} Yamauchi S., Koyama K., 1993, ApJ, 404,
620

\bibitem[\protect\citeauthoryear{Yamauchi et
al.}{1996}]{yamauchi96} Yamauchi S., Kaneda H., Koyama K.,
Makishima K., Matsuzaki K., Sonobe T., Tanaka Y., Yamasaki N., 1996, PASJ,
48, L15

\end{thebibliography}
\end{document}